\documentclass[fdp,fleqn]{w-art}
\usepackage{times}
\usepackage{w-thm}
\usepackage[]{graphicx}

\begin{document}

\DOIsuffix{theDOIsuffix}
\Volume{xx}
\Month{xx}
\Year{2008}

\pagespan{1}{}



\title[Is dark energy an effect of averaging?]{Is dark energy an effect of averaging?}

\author{Nan Li, Marina Seikel and Dominik J. Schwarz
\footnote{Talk at ``Balkan workshop 2007", Kladovo (Serbia).}
}
\address{Fakult\"at f\"ur Physik, Universit\"at Bielefeld, Postfach 100131, D-33501 Bielefeld}
\date{\today}

\begin{abstract}
The present standard model of cosmology states that the known
particles carry only a tiny fraction of total mass and energy of the
Universe. Rather, unknown dark matter and dark energy are the
dominant contributions to the cosmic energy budget. We review the
logic that leads to the postulated dark energy and present an
alternative point of view, in which the puzzle may be solved by
properly taking into account the influence of cosmic structures on
global observables. We illustrate the effect of averaging on the
measurement of the Hubble constant.
\end{abstract}

\maketitle

\section{Introduction}

The concordance model of cosmology states that only $4\%$ of the
mass/energy content of the Universe is carried by identified
particles, and the dominant contributions are unknown dark matter
($22\%$) and mysterious dark energy ($74\%$) \cite{Spergel}. It is
said that we are living in the age of ``precision cosmology'', which
is justified as a statement on the precision of observations, but
certainly not on their understandings. The purpose of this talk is
to recapitulate the genesis of the dark energy hypothesis and to
show that it poses a serious crisis to theoretical cosmology. We
argue that a possible resolution for this crisis may come from the
proper treatment of effects from averaging.

\section{Dark energy}

The vast majority of cosmic background photons (photons without
identified source) are due to the cosmic microwave background (CMB)
radiation, which is isotropic (apart from a component due to our
proper motion) with fluctuations at the level of $10^{-5}$. Thus the isotropy of the
Universe (around us) is an observational fact. On the other hand,
there is no evidence that the Milky Way would be a distinguished
place in the Universe and thus it seems reasonable to follow
Copernicus to assume that we are not living at the centre of the
Universe. Therefore, we end up with a homogeneous and isotropic
model. The most general line element in such a
Friedmann-Lema\^{\i}tre model is given by
\[
{\rm d}s^2 = - {\rm d}t^2 + a^2\left({{\rm d}r^2\over 1 - K r^2} + r^2
{\rm d}\Omega^2\right),
\]
where $a= a(t)$ denotes the scale factor and $K/a^2$ the spatial
curvature ($K = -1,0,+1$) of the Universe. In this model physical
distances evolve with time, $r_{\rm ph} = a f(r)$ (we may fix the
present scale factor $a_0\equiv1$), where 
$f(r) = \{{\rm Arsinh}(r), r, \arcsin(r)\}$ for $K= \{-1,0,1\}$. 
It is useful to define the rate
of linear expansion $H \equiv \dot{a}/a$. Without using Einstein's
equations, just assuming that photons travel on null geodesics, one
finds that light from distant objects is red-shifted by $z\equiv
\Delta\lambda/\lambda = 1/a -1$ and the so-called luminosity
distance (defined by the observed flux from a source with known
luminosity $d_{\rm L} = \sqrt{L/4 \pi F}$) satisfies the Hubble law
\[
d_{\rm L} = \frac{c}{H_0}\left[ z + \left(1-q_0\right) \frac{z^2}{2} + {\cal O}(z^3) \right]
\]
at small redshifts $z \ll 1$. Observations show that $H_0 \approx
70$ km/s/Mpc $> 0$ and thus we know that the Universe expands. The
typical time scale of the cosmos is estimated as $t_{\rm H} \approx
14$ Gyr, in concordance with the estimate on the age of the oldest
objects in the Universe \cite{oldstar}. Another interesting
kinematic quantity is the acceleration (deceleration) of the cosmic
expansion, encoded by the deceleration parameter $q \equiv -
(\ddot{a}/a)/H^2$. Measurements of supernovae of type Ia (SN Ia)
indicate that $q_0 < 0$.

One might wonder what could be the reason for the high symmetry of
the observed Universe. The answer of modern cosmology is the idea of
cosmological inflation \cite{inflation}, an epoch in the very early
Universe that is held responsible for the Universe's homogeneity and
isotropy. On top of that, generic predictions of cosmological
inflation are spatial flatness ($\Omega_{\rm K}=0$) and the
existence of tiny fluctuations \cite{Chibisov}, as observed at a
level of $10^{-5}$ in the CMB.

We may ask how strong is the evidence for the acceleration of the
Universe, even before we wish to invoke a model for the matter
content, without assuming the validity of Einstein's equations. This
test goes as follows. We ask at what level of confidence we can
reject the null hypothesis that the Universe never accelerated (at
$z <$ few). For a spatially flat model, as suggested by the idea of
cosmological inflation, our null hypothesis can be expressed as
\[
d_{\rm L} < \frac{c}{H_0} (1+z)\ln (1+z).
\]
Based on two different SN Ia data sets, two different fitting
methods and two different calibration methods, we conclude that the
null hypothesis is rejected at $> 5 \sigma$ \cite{Seikel}.

\begin{figure}
\begin{center}
\includegraphics*[width=8cm]{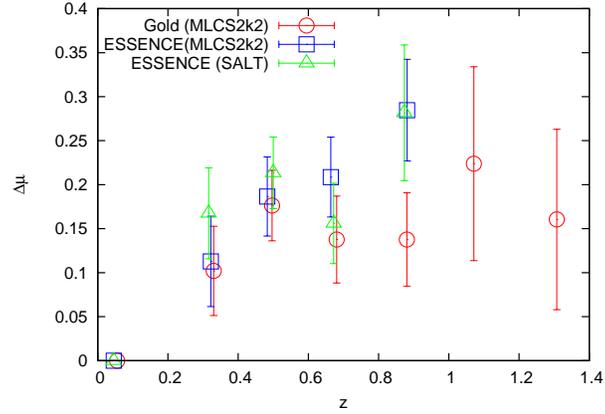}
\end{center}
\caption{\label{deltamu} The binned magnitudes $\Delta\mu$ (see text) for different SN data sets and
light-curve fitting methods.}
\end{figure}

This test can also be done in a calibration-free way, i.e. without
assuming certain values for the Hubble constant $H_0$ and the
luminosity of the SNe. Instead, we consider the averaged value of
the apparent magnitude $m$ of SNe within a certain redshift bin
relative to the averaged apparent magnitude $m_{\rm{nearby}}$ of
nearby SNe with $z<0.2$. Defining a new quantity
\[
\Delta\mu = m - m_{\rm{nearby}} = 5\log_{10} d_{\rm L} - 5\log_{10}
d_{\rm{L,nearby}},
\]
the null hypothesis can be written as $\Delta\mu < 0$. For the test,
we average $m$ within redshift bins of width 0.2. In Fig.
\ref{deltamu}, the result is shown for two data sets (Gold
\cite{Riess07} and ESSENCE \cite{Wood}) and two fitting methods
(MLCS2k2 \cite{Jha} and SALT \cite{Guy}). The evidence for
acceleration can be obtained by dividing $\Delta\mu$ by its error.
The result for each redshift bin is given in Table \ref{evidence}.
Also given is the evidence obtained from averaging over the apparent
magnitude of all SNe with redshift $z\ge 0.2$. It is noticeable that
using the calibration-free test weakens the evidence for
acceleration (4.3$\sigma$) compared to the result of the previous
test (5.2$\sigma$).

\begin{table}
\begin{center}
\begin{tabular}{@{}cccc}
\hline
           & Gold 2007 & ESSENCE  & ESSENCE \\
$z$        & (MLCS2k2) & (MLCS2k2)& (SALT)  \\
\hline
0.2 -- 0.4 & 2.0       & 2.2      & 3.2     \\
0.4 -- 0.6 & 4.4       & 4.2      & 5.2     \\
0.6 -- 0.8 & 2.8       & 4.6      & 3.4     \\
0.8 -- 1.0 & 2.6       & 4.9      & 3.7     \\
1.0 -- 1.2 & 2.0       &          &         \\
1.2 -- 1.4 & 1.6       &          &         \\
\hline
$\ge$ 0.2  & 4.3       & 5.2      & 5.6     \\
\hline
\end{tabular}
\end{center}
\caption{\label{evidence} Statistical evidence for acceleration from
different redshift bins for different SN data sets and light-curve fitting
methods.}
\end{table}

Let us now ask what that implies for the dynamical model of the
Universe. The most general equation of motion for a homogeneous and
isotropic Universe under the assumption of diffeomorphism invariance
and at most second order derivatives on the metric (Lovelock theorem
\cite{lovelock}) provides us with the Friedmann equation and the
continuity equation ($c=1$)
\[
H^2 + \frac{K}{a^2} - \Lambda = \frac{8\pi G}{3} \epsilon,
\qquad \dot{\epsilon} + 3 H (\epsilon + p) = 0.
\]
$\epsilon$ and $p$ denote the energy density and pressure. $G$ and
$\Lambda$ are Newton's constant and the cosmological constant. In
order to close this system of equations, we also need to impose an
equation of state $p = p(\epsilon)$, which in many cases can be
expressed as $w = p/\epsilon$. A dimensionless energy density
$\Omega \equiv \epsilon/(3 H^2/8 \pi G)$ is introduced for
convenience.

We have already mentioned that cosmological inflation predicts a
vanishing spatial curvature for all practical purposes and thus we
can put $K=0$. In the present Universe, matter is non-relativistic
(cold), which means $p \ll \epsilon$ and thus we put $p=0$. If we
now would put $\Lambda =0$ as well, we arrive at the Einstein-de
Sitter Universe, which predicts $q(t) = 1/2$ and $t_0 = 2 t_{\rm
H}/3 \approx 9$ Gyr. Therefore, we cannot stick to this simple
model, as we have seen that $q < 0$ at high confidence level and the
oldest stars are known to be as old as $12$ Gyr or even older
\cite{oldstar}.

These facts became clear only about ten years ago --- the 1998
cosmology revolution. One can rewrite the dynamic equations into
\[
-3\frac{\ddot{a}}{a} = 4\pi G( \epsilon + 3 p) - \Lambda < 0.
 \]
The inequality and thus acceleration, can occur if either $\Lambda
> 4\pi G \epsilon$ (equivalent to $\Omega_\Lambda > \Omega_{\rm
m}/2$) or there exists some form of energy density with $p < -
\epsilon/3$, a form of energy that has never been observed in the
laboratory so far and is thus dubbed the ``dark energy''. For $p = -
\epsilon$ this reproduces the behaviour of the cosmological constant
(as $\dot{\epsilon} = 0$ in that case), which is thus a special case
of dark energy.

So a minimal model consistent with the predictions of cosmological
inflation and allowing for the observed accelerated expansion of the
Universe is given by $K= p = 0$ and $\Omega_\Lambda > \Omega_{\rm
m}/2 = (1 - \Omega_\Lambda)/2$ and thus $\Omega_\Lambda
> 1/3$ and $\Omega_{\rm m} < 2/3$. This model is called the flat
$\Lambda$CDM (cold dark matter) model or concordance model, as it is
currently in good agreement with all cosmological observations.

One can test for a possible deviation from $K=0$ or $p=0$ by means
of CMB observations, SN Ia and the baryon acoustic oscillations and
finds that there is no observational evidence to question the
validity of the flat $\Lambda$CDM model \cite{Spergel,Davis}.

Let us stress that an important assumption in the reasoning described above is the
homogeneity and isotropy of the Universe and thus it is also
important to test this assumption with the help of SN Ia. A simple
test can be done by fitting the Hubble diagram ($d_{\rm L}$ vs. $z$)
for SNe from just one hemisphere and comparing the result of the fit
to the opposite half of the sky \cite{Weinhorst}. Weinhorst and one of the authors
have analysed four different SN data sets and find a statistically significant
deviation from the isotropy of the Hubble diagram. However, it seems
that these anisotropies are aligned with the equatorial
coordinate system (aligned with the Earth's axis of rotation) and
are thus likely to reflect systematic problems in the SN Ia
observation or analysis. Another recent analysis revealed anisotropies in the Hubble diagram
obtained by the HST Key Project \cite{McClure}. If these anisotropies would be confirmed by
future SN surveys and a systematic effect could be excluded, this
would pose a serious challenge to the interpretation of Hubble
diagrams in the Friedmann-Lema\^{\i}tre context.

\section{Problems of the concordance model}

Thus it seems that the concordance model is able to describe
everything, however there are three conceptual problems in this
model.

{\bf The cosmological constant problem.} We miss out a fundamental
understanding of the vacuum energy of quantum fields. This vacuum
energy density is indistinguishable from a cosmological constant and
thus we have introduced an element to the Universe that we do not
understand. The solution for this issue must be expected from a
future theory of quantum gravity, as gravity (i.e. geometry of
space-time) is the only ``force'' that couples to the energy density
of the (quantum field) vacuum. Thus this problem is obviously beyond
the scope of what we can hope to understand currently.

{\bf The coincidence problem(s).} The second problem however is a
pure cosmological one. It is called the coincidence problem and
there are two ways to formulate it. The first formulation has some
anthropic touch, but is widely used: how does it come that
$\Omega_{\rm m}(t_0) \sim \Omega_\Lambda(t_0)$? This is indeed
surprising. Assume we would have lived when the Universe had half
(twice) of its present linear size [at $z = 1$ (-1/2)], the ratio
$\Omega_{\rm m}/\Omega_\Lambda$ would be larger (smaller) by almost
an order of magnitude. It is therefore quite surprising that we
observe $\Omega_\Lambda \approx 0.7 > 1/3$ at $t_0$.

One can reformulate the above situation via replacing mankind by
large-scale-structures on the typical scale for structure formation.
According to inflationary cosmology, the primordial fluctuations in
matter and metric are free of any scale. The only scale that enters
in the physics of the large scale structure is the horizon scale at
matter-radiation equality. This scale appears because the radiation
pressure acts against the gravitational instability in the early
Universe and structure growth starts only once the pressure support
fades away. Thus the shape of the power spectrum of density
fluctuations is changing at this scale. According to the concordance
model, the horizon scale at matter-radiation equality is seen at the
$100$ Mpc scale today. Scales above that show the primordial slope
of the spectrum, scales below that exhibit a substantially decreased
slope. It is now interesting to observe that those structures grow
to typical density contrasts of order $(\delta
\epsilon/\epsilon)_{z=0.x} \sim (1+ z_{\rm eq}) (\delta
\epsilon/\epsilon)_{\rm eq} \sim 0.1$ and it turns out that $z =
{\cal O}(0.1)$ is the epoch when the typical scale of hierarchical
structure formation starts to evolve non-linear (i.e. it decouples
from the overall cosmological expansion). The coincidence is that
this is also the epoch when acceleration seems to set in \cite{Schwarz}. Note that
these large structures are indeed observed, e.g. the Sloan great
wall and various big voids, as big as a few $100$ Mpc
\cite{largest_structures}.

{\bf The averaging problem.} This is an old question in general
relativity \cite{averaging}. Solving the Einstein equations for an
averaged space-time is certainly not the same as averaging the
Einstein equations for a very inhomogeneous and anisotropic
space-time, because these two operations, i.e. averaging and time
evolution do not commute. Thus the issue is, how big is this effect
and what are its observational signatures. Again the scale of
interest must be the $100$ Mpc scale, as on much smaller scales, say
$1$ Mpc objects like galaxies have already presumably undergone the
process of violent dynamical relaxation and formed stationary bound
systems that cannot be treated like dust particles anyway. However
it is clear that for objects like the Sloan great wall, which just
started to collapse yesterday, there might be important effects
showing up in the averaging process.

In summary, these three aspects are entangled with the issues of
inhomogeneities and anisotropies in the local Universe. The
beginning of the domination of dark energy coincides with the onset
of structure formation. Consequently, light may be shed on the dark
energy crisis (i.e. the accelerated expansion of the Universe) by
studying the averaging problem in the perturbed space-time.

An attempt for setting up an formalism to handle this problem has
been pioneered by Buchert \cite{Buchert:1999er,Buchert:2007}.

\section{Cosmological backreaction}

These problems provide a good motivation to consider the possibility
that dark energy is not a fundamental component of the Universe, but
our averaged description of the Universe is not appropriate for a
correct interpretation of observations \cite{Schwarz,backreaction,Kolb:2004am}. 
Indeed many cosmological
observations are averages. Two important examples are the power
spectrum $P(k)$, which is a Fourier transform and thus a volume
average weighted by a factor $\exp (i{\bf k}\cdot{\bf x})$, and the
Hubble constant $H_0$. Let us consider an idealised measurement of
$H_0$ \cite{Tully:2007ue}. One picks $N$ standard candles in a local
physical volume $V$ (e.g. SN Ia in Milky Way's neighbourhood out to
$\sim100$ Mpc), measures their luminosity distances $d_i$ and
recession velocities $v_i=cz_i$, and takes the average $H_0 \equiv
\frac{1}{N}\sum_{i=1}^N \frac{v_i}{d_i}$. In the limit of a very big
sample, it turns into a volume average $H_0 = \frac{1}{V}\int
\frac{v}{d} {\rm d}V$. For objects at $z \ll 1$, the spatial average
is appropriate for the average over the past light cone, as the
expansion rate of the Universe is not changing significantly at time
scales much shorter than the Hubble time.

We now recapitulate the formulation of Buchert, which we will use
below to argue that the averaging effect is indeed sizeable and does
give important contributions to cosmology:

Buchert's setup is well adapted to the situation of a real observer.
On large scales, a real observer is comoving with matter, uses her
own clock, and regards space to be time-orthogonal. These conditions
are the definition of the comoving synchronous coordinate system.
Buchert used physically comoving boundaries, which is the most
natural approach in this setup. In the following the Universe is
assumed to be irrotational as a consequence from cosmological
inflation.

In synchronous coordinates, the metric of the inhomogeneous and
anisotropic Universe is $\mbox{d}s^2 = -\mbox{d}t^2 + g_{ij}(t,{\bf
x})\mbox{d}x^i\mbox{d}x^j$, and the spatial average of an observable
$O(t,\bf x)$ in a physically comoving domain $D$ at time $t$ is
defined as \cite{Buchert:1999er}
\begin{eqnarray}
\langle O \rangle_D\equiv \frac{1}{V_D(t)}\int_D O(t,{\bf x})
\sqrt{\mbox{det}g_{ij}}\mbox{d}\bf x,\label{average}
\end{eqnarray}
where $V_D(t) \equiv \int_D \sqrt{\mbox{det}g_{ij}} \mbox{d}{\bf x}$
is the volume of the comoving domain $D$. We may introduce an
effective scale factor $a_D$ \cite{Buchert:1999er}
\begin{eqnarray}
\frac{a_D}{a_{D_0}}\equiv
\left(\frac{V_D}{V_{D_0}}\right)^{1/3}. \label{ad}
\end{eqnarray} 
The effective Hubble
expansion rate is thus defined as $H_D\equiv \dot{a}_D/a_D=\langle
\theta\rangle_D/3$ ($\theta$ being the volume expansion rate).

From the definition (\ref{average}), we obtain effective Friedmann
equations \cite{Buchert:1999er} from averaging Einstein's equations,
\begin{eqnarray}
\left(\frac{\dot{a}_D}{a_D}\right)^2= \frac{8\pi G}{3}\epsilon_{\rm
eff}, \qquad -\frac{\ddot{a}_D}{a_D} = \frac{4\pi
G}{3}(\epsilon_{\rm eff}+3p_{\rm eff}), \label{buchert}
\end{eqnarray}
where $\epsilon_{\rm eff}$ and $p_{\rm eff}$ are the energy density
and pressure of an effective fluid,
\begin{eqnarray}
\epsilon_{\rm eff}\equiv\langle\epsilon\rangle_D-\frac{1}{16\pi
G}\left(\langle Q\rangle_D+\langle R \rangle_D\right), \qquad p_{\rm
eff}\equiv-\frac{1}{16\pi G}\left(\langle Q\rangle_D-
\frac{1}{3}\langle R\rangle_D\right). \label{rhoeff}
\end{eqnarray}
$\langle Q\rangle_D\equiv\frac{2}{3}(\langle
\theta^2\rangle_D-\langle \theta\rangle_D^2)-2\langle
\sigma^2\rangle_D$ is the kinematical backreaction ($\sigma^2$ being
the shear scalar), and $\langle R\rangle_D$ the averaged spatial
curvature. They are related by the integrability condition
\cite{Buchert:1999er}
\begin{eqnarray}
(a_D^6\langle Q\rangle_D)^{^{\textbf{.}}}+a_D^4(a_D^2\langle
R\rangle_D)^{^{\textbf{.}}}=0.\label{integ}
\end{eqnarray}
We define the equation
of state for the effective fluid as $w_{\rm eff}\equiv p_{\rm
eff}/\epsilon_{\rm eff}$. It is highly remarkable that any spatially
averaged dust model can be described by an effective
Friedman-Lema\^itre model.

We can map this effective fluid on a model with dust and ``dark
energy". Let $n$ be the number density of dust particles, and $m$ be
their mass. For any comoving domain $\langle n\rangle_D=\langle
n\rangle_{D_0}(a_{D_0}/a_D)^3$. For a dust Universe, in which
$\epsilon(t, {\bf x}) \equiv mn(t, {\bf x})$, we identify
$\epsilon_{\rm m} \equiv \langle \epsilon\rangle_{D}= m \langle
n\rangle_{D}$, and from (\ref{rhoeff}) the dark energy is hence
described by $\epsilon_{\mathrm{de}}=-(\langle Q\rangle_D+\langle
R\rangle_D)/(16\pi G)$. From (\ref{integ}) we find that constant
$\langle Q\rangle_D=-\langle R\rangle_D/3$ corresponds to the case
of a cosmological constant $\Lambda=\langle Q\rangle_D$. Equations
(\ref{buchert}) and (\ref{rhoeff}) are not closed and additional
input is required. Below we close them by means of cosmological
perturbation theory.

To study the scale dependence of physical observables $\langle
Q\rangle_D$, $\langle R\rangle_D$, $\langle \epsilon\rangle_D$,
$H_D$ and $w_{\mathrm{eff}}$, we calculate them to second order in a
perturbative series of the effective scale factor $a_D$. We start
from a spatially flat dust model. Its scale factor $a(t)$ is
different from the effective scale factor $a_D$ in (\ref{ad}), and
their relation was provided in \cite{Li:2007ci}. In synchronous
gauge, the linear perturbed metric is
\begin{eqnarray}
\mbox{d}s^2=-\mbox{d}t^2+a^2(t)[(1-2\Psi)\delta_{ij}+D_{ij}\chi]\mbox{d}x^i
\mbox{d}x^j,\nonumber
\end{eqnarray}
where $\Psi$ and $\chi$ are the scalar metric perturbations, $D_{ij}
\equiv \partial_{i}\partial_{j}-\frac{1}{3}\delta_{ij}\Delta$ and
$\Delta$ denotes the Laplace operator in three-dimensional Euclidean
space. The solutions for $\Psi$ and $\chi$ are given in terms of the
time independent peculiar gravitational potential $\varphi({\bf
x})$: $\Psi=\frac{1}{2}\Delta \varphi
t_0^{4/3}t^{2/3}+\frac{5}{3}\varphi$ and $\chi=-3\varphi
t_0^{4/3}t^{2/3}$ (only growing modes are considered)
\cite{Li:2007ci,Kolb:2004am}. $\varphi$ is related to the
hypersurface-invariant variable $\zeta$ \cite{Bardeen:1988hy} by
$\zeta=\frac{1}{2}\Delta \varphi
t_0^{4/3}t^{2/3}-\frac{5}{3}\varphi$.

Following \cite{Li:2007ci}, with the help of the integrability
condition, we yield the scale dependence of the averaged physical
observables up to second order \cite{Li:2007ny} ($\langle
O\rangle_{D1}\equiv \int_D O {\rm d}{\bf x}/\int_D {\rm d}{\bf x}$
hereafter)
\begin{eqnarray}
\langle Q\rangle_D&=&\frac{a_{D_0}}{a_D}B(\varphi)t^2_0,\label{q2}\\
\langle R\rangle_D&=&\frac{20}{3}\frac{a_{D_0}^2}{a_D^2}\langle
\Delta \varphi\rangle_{D1}-5\frac{a_{D_0}}{a_D}B(\varphi)t^2_0,\label{r2}\\
\langle \epsilon\rangle_D&=&\frac{1}{6\pi G t_0^2}\frac{a_{D_0}^3}{a_D^3},\label{rho2}\\
H_D&=&\frac{2}{3t_0}\frac{a_{D_0}^{3/2}}{a_D^{3/2}}
\left[1-\frac{5}{4}\frac{a_D}{a_{D_0}}t_0^2\langle \Delta
\varphi\rangle_{D1}+\frac{3}{4}\frac{a_D^2}{a_{D_0}^2}t_0^4\left(B(\varphi)
-\frac{25}{24}\langle \Delta \varphi\rangle_{D1}^2\right)\right],\label{theta2}\\
w_{\rm eff}&=&\frac{5}{6}\frac{a_D}{a_{D_0}}t_0^2\langle\Delta
\varphi\rangle_{D1}-\frac{a_D^2}{a_{D_0}^2}t_0^4\left(B(\varphi)-\frac{25}{12}\langle
\Delta \varphi\rangle_{D1}^2\right),\label{w2}
\end{eqnarray}
with $B(\varphi)\equiv \langle
\partial^i(\partial_i\varphi\Delta \varphi)-
\partial^i(\partial_j\varphi\partial^j\partial_i\varphi)\rangle_{D1}
-\frac{2}{3}\langle \Delta \varphi\rangle_{D1}^2$. We see from
(\ref{q2}) -- (\ref{w2}) that these quantities are functions of
surface terms only, so all their information is encoded on the
boundaries. The dependence on cosmic time of these averaged
quantities can be found in \cite{Li:2007ci}, and their leading terms
are gauge invariant \cite{Li:2007ci,Bruni:1996im}.

We now turn to the estimate of the effects of cosmological
backreaction as a function of the averaging scale $r\sim
V_{D_{0}}^{1/3}$. We show below that cosmological averaging produces
reliable and important modifications to local physical observables
and determine the averaging scale at which $10\%$ corrections show
up.

Effective acceleration of the averaged Universe occurs if
$\epsilon_{\mathrm{eff}}+3p_{\mathrm{eff}}<0$, i.e. $\langle
Q\rangle_D>4\pi G\langle \epsilon\rangle_D$. Thus we estimate
\begin{eqnarray}
\left|\frac{\langle Q\rangle_D}{4\pi G\langle
\epsilon\rangle_D}\right|=
\frac{3}{2}\frac{a_D^2}{a_{D_0}^2}B(\varphi)t_0^4 = \frac{8}{27}
\frac{R_{\rm H}^4}{(1+z)^2}
B(\varphi)\sim\frac{8}{75}\frac{1}{(1+z)^2} \left(\frac{R_{\rm
H}}{r}\right)^4 \mathcal{P}_{\zeta}. \label{onsetq}
\end{eqnarray}
$R_{\rm H}=3.00\times 10^3h^{-1}$ Mpc is the present Hubble distance
and $\mathcal{P}_{\zeta}=2.35\times 10^{-9}$ the dimensionless power
spectrum \cite{Spergel} ($\zeta\approx-5\varphi/3$ on superhorizon
scales). In (\ref{onsetq}), we use $a=1/(1+z)$ and $t_0=2R_{\rm
H}/3$, since $B$ is already a second order term. Each derivative in
$B$ is estimated as a factor of $1/r$ in front of the power
spectrum. As $\varphi$ is constant in time, we identify today's
$\mathcal{P}_{\varphi}$ with $9\mathcal{P}_{\zeta}/25$ at
superhorizon scales. If $|\langle Q\rangle_D/ 4\pi G\langle
\epsilon\rangle_D| > 0.1$, cosmological backreaction must be taken
into account seriously. From (\ref{onsetq}), we find
\begin{eqnarray}
r_Q < \frac{21 h^{-1}~{\rm Mpc}}{\sqrt{1+z}}. \nonumber
\end{eqnarray}
For observations at $z\ll 1$, $r_Q < 30$ Mpc ($h = 0.7$).

The criterion for the scale at which the effect of the averaged
spatial curvature $\langle R\rangle_D$ emerges is determined by
\begin{equation}
\left|\frac{\epsilon_{\rm eff}}{\langle \epsilon \rangle_D}-1\right|
\approx \left|\frac{\langle R\rangle_D}{16\pi G\langle
\epsilon\rangle_D}\right| \sim
\frac{2}{3}\frac{1}{1+z}\left(\frac{R_{\rm H}}{r}\right)^2
\sqrt{\mathcal{P}_{\zeta}}.\nonumber
\end{equation}
We expect a $10\%$ effect from $\langle R\rangle_D$ within
\begin{eqnarray}
r_R < \frac{54h^{-1}~{\rm Mpc}}{\sqrt{1+z}}. \nonumber
\end{eqnarray}
At small redshifts, $r_R < 77$ Mpc. Moreover, a $1\%$ effect is
expected up to scales of $\sim 240$ Mpc. Note that the curvature of
the Universe has been measured at the few per cent accuracy in the
CMB \cite{Spergel}. It was shown in \cite{Clarkson} that even small
curvatures might influence the analysis of high-$z$ SNe
significantly.

\section{Signature of cosmological backreaction: the effective Hubble rate}

It is now interesting to discuss the fluctuation of the local
measurement of the Hubble expansion rate,
\begin{eqnarray}
\left|\frac{H_D-H_0}{H_0}\right|
&\sim&\frac{1}{3}\frac{1}{1+z}\left(\frac{R_{\rm H}}{r}\right)^2
\sqrt{\mathcal{P}_{\zeta}}.\label{onseth}
\end{eqnarray}
The leading contribution to this effect involves the averaged
curvature terms and therefore cannot be thoroughly described in a
Newtonian setup \cite{NewtonianHubble}. An effect
larger than $10\%$ shows up for
\begin{eqnarray}
r_H < \frac{38h^{-1}~{\rm Mpc}}{\sqrt{1+z}},\nonumber
\end{eqnarray}
which reads $r_H < 54$ Mpc at small redshifts.

\begin{figure*}
\begin{center}
\includegraphics[width=0.3\linewidth,angle=270]{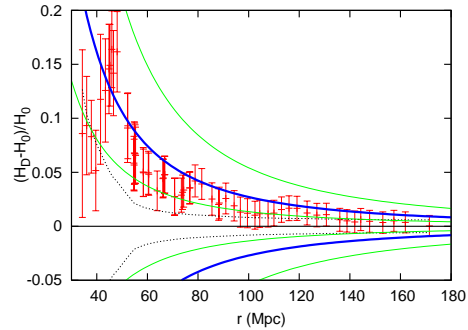}
\end{center}
\caption{\label{fig1} The scale dependence of the normalised
difference between the effective Hubble rate $H_D$ and its
``global'' value $H_0=72$ km/s/Mpc (Data from the HST Key Project
\cite{Freedman:2000cf}). We expect that all data points should lie
around the two thick lines ($\sim \pm 1/r^2$) given in
(\ref{onseth}). Thin lines indicate the theoretical error of our
estimate. Dashed lines are the expected statistical noise for a
perfectly homogeneous and isotropic model.}
\end{figure*}

Let us compare (\ref{onseth}) with observations from the HST Key
Project \cite{Freedman:2000cf}. We use 68 individual measurements of
$H_0$ in the CMB rest frame from SN Ia, the Tully-Fisher and
fundamental plane relations (Tables 6, 7 and 9 in
\cite{Freedman:2000cf}). The respective objects lie at distances
between 14 Mpc and 467 Mpc. As (\ref{onseth}) can be trusted only
above 30 Mpc, we drop the 4 nearest objects. Be $r_i$, $H_i$ and
$\sigma_i$ the distance, Hubble rate and its $1\sigma$ error for the
$i'$th datum, with increasing distance. We calculate the averaged
distance for the nearest $k$ objects by $\bar{r}_k= \sum_{i=1}^k g_i
r_i / \sum_{i=1}^k g_i$, with weights $g_i = 1/\sigma_i^2$. An
analogue holds for the averaged Hubble rate $\bar{H}_k$, i.e. $H_D$
for different subsets. The empirical variance of each subset is
$\bar{\sigma}^2_k=\sum_{i=1}^k g_i
(H_i-\bar{H}_k)^2/[(k-1)\sum_{i=1}^k g_i]$. We stress that
(\ref{onseth}) is insensitive to global calibration issues.

The comparison of our result (\ref{onseth}) with the HST Key Project
data is shown in Fig. 2. The ``global" $H_0$ is the central HST
value 72 km/s/Mpc \cite{Freedman:2000cf}. We find that the
theoretical curve ($\propto 1/r^2$) matches the experimental data
fairly well, without any fit parameter in the panel. Since
(\ref{onseth}) is just an estimate, we vary the result by factors of
$0.5$ and $2$, which gives lower and upper estimates to the
theoretical curve. In Fig. 2, the experimental data and their error
bars are in agreement with these estimates.

Before we can claim that we observe the expected $1/r^2$ behaviour
of (\ref{onseth}) and thus evidence for cosmological backreaction,
we have to make sure that the statistical noise cannot account for
it. The expected noise is given by means of the variance of
(\ref{onseth}) in the situation of a perfectly homogeneous and
isotropic model with measurement uncertainties $\sigma_i$. In the
case of a perfectly homogenous coverage of the averaged domain with
standard candles, we expect a $1/r^{3/2}$ behaviour. In Fig. 2, we
show the statistical noise for the actual data set, which turns out
to be well below our estimate (\ref{onseth}) and the data points. At
$\sim 50$ $(80)$ Mpc, the value of $H_D$ differs from its global one
by about $10\%$ $(5\%)$, whereas the expected variance for a
homogeneous and isotropic Universe is only $3\%$ $(1\%)$.

\section{Toward a new model and open issues}

What can we conclude so far? We showed that the large scale structure affects the observed Hubble expansion rate and that the averaged spatial curvature leads to sizeable corrections up to scales of about 200 Mpc. Consequently, we must expect that also other cosmological probes at $z<1$, like 
galaxy clusters or galaxy redshift surveys, could be affected by cosmic averaging. 

Is that enough to explain dark energy? Currently, this is not the case. 
In our framework an effective acceleration is possible below scales of 20 Mpc. But SN Ia observations 
show strong evidence for the acceleration of the Universe, based on objects much more distant than that. It seems to take more to explain dark energy. A quite attractive idea is a Universe in which 
large voids dominate the Universe's volume, see \cite{Wiltshire, Kolb, Notari, Mattsson}. These works
rely on well motivated toy models. It remains to be shown that these void dominated cosmologies
follow from the inflationary paradigm.  

On the other hand, our analysis is based on an almost scale-invariant power spectrum predicted by inflationary cosmology, but is limited to $z \ll 1$, as the average must be taken on the past 
light cone and not on a spatial slice (see, e.g.~\cite{lightcone}). A formalism (like Buchert's one for 
spatial averaging) for the light-cone averaging of an arbitrary inhomogeneous and anisotropic space-time is needed before we can know if dark energy is merely an illusion due to averaging. 

\section*{Acknowledgements}
We thank Adam Riess for suggesting the calibration independent test presented in figure 1 and table 1.
DJS would like to thank the organisers of the ``Balkan workshop 2007'' for having invited him to such a lively and inspiring meeting. DJS would also like to thank  SEENET-MP and the DFG for support of his travel. We are grateful to the DFG to support our research under grant GKR 881.

\end{document}